\documentclass[twocolumn]{aastex61}
\usepackage{graphicx}
\usepackage{subfigure}
\usepackage{mathscinet}

\received{July 3, 2017}
\revised{September 6, 2017}
\accepted{November 6, 2017}

\submitjournal{AAS Journals}

\shorttitle{Activity of TRAPPIST-1}
\shortauthors{Roettenbacher \& Kane}

\begin{document}

\title{The Stellar Activity of TRAPPIST-1 and Consequences for the Planetary Atmospheres  }

\correspondingauthor{Rachael M.\ Roettenbacher}
\email{rachael@astro.su.se}

\author[0000-0002-9288-3482]{Rachael M.\ Roettenbacher}
\affiliation{Department of Astronomy\\
Stockholm University \\
SE-106 91 Stockholm, Sweden}

\author[0000-0002-7084-0529]{Stephen R.\ Kane}
\affiliation{Department of Physics \& Astronomy \\
San Francisco State University \\
San Francisco, CA 94132, USA}

\begin{abstract}

The signatures of planets hosted by M dwarfs are more readily detected with transit photometry and radial velocity methods than those of planets around larger stars.  Recently, transit photometry was used to discover seven planets orbiting the late-M dwarf TRAPPIST-1.  Three of TRAPPIST-1's planets fall in the Habitable Zone, a region where liquid water could exist on the planetary surface given appropriate planetary conditions. 
We aim to investigate the habitability of the TRAPPIST-1 planets by studying the star's activity and its effect on the planets.  
We analyze previously-published space- and ground-based light curves and show the photometrically-determined rotation period of TRAPPIST-1 appears to vary over time due to  complicated, evolving surface activity.  The dramatic changes of the surface of TRAPPIST-1 suggest that rotation periods determined photometrically may not be reliable for this and similarly active stars.
While the activity of the star is low, we use the premise of the ``cosmic shoreline'' to provide evidence that the TRAPPIST-1 environment has potentially led to the erosion of possible planetary atmospheres by extreme ultraviolet stellar emission.

\end{abstract}

\keywords{stars:  variables:  general --- stars: activity --- starspots --- stars:  individual (TRAPPIST-1)}

\section{Introduction} \label{sec:intro}

Late-M dwarfs are among the most active stars, exhibiting photometric variations indicative of starspots and flares.  Starspots are cool, dark regions of the stellar surface that manifest from strong magnetic fields stifling efficient energy transport in the outer convective layers \citep[e.g.,][]{str09}.  Starspots are often detected as photometric variations as they rotate in and out of view.  Starspots are frequently used to determine the star's rotation period and can additionally be used to map the stellar surface \citep[e.g.,][]{roe16a}.  Flares, characterized by rapid increases in optical and X-ray flux, are energetic events driven by the reconnection of magnetic fields \citep[e.g.,][and references therein]{pri02}.
Late-M dwarf stellar flares, for example, have been observed to have energies in excess of $10^{37}$ erg \citep{wu15}.  While not every flare has such high energy, some M dwarfs have a flare frequency of several flares per day \citep[e.g., GJ 1243;][]{dav16}.  Frequent flares, even weak ones, will affect the regions close to the stars; for M dwarfs, this includes the so-called Habitable Zone (HZ), a region at a distance from the star in which water on the planetary surface would be liquid \citep{kas93,kan12,kop13,kop14,kas14}.

TRAPPIST-1 (2MASS J23062928-0502285) is an M8 dwarf \citep{lie06} that has recently been determined to host at least seven terrestrial-like planets \citep{gil17}.  The host star has a radius of $R = 0.117 \pm 0.004 ~\ R_\odot$, mass of $M = 0.080 \pm 0.008~\ M_\odot$, and luminosity of $L = 0.00052 \pm 0.00004 ~\ L_\odot$ \citep{gil16}. The stellar parameters of TRAPPIST-1 reported by \citet{gil17} suggest that the e, f, and g planets are in the HZ. Based on their TRAPPIST-South ground-based photometry, \citet{gil16} characterized TRAPPIST-1 as a moderately active star. The nearly-continuous, 20-day \emph{Spitzer} light curve of \citet{gil17} contained only two flares at BJD = 2457659.38 and 2457667.13 (time of maximum flare brightness).  However, the preliminary \emph{K2} light curve shows flares of higher frequency \citep{vid17}. 

Here, we investigate the activity of TRAPPIST-1 and its implications for the planetary atmospheres and subsequent habitability. In Section \ref{sec:obs}, we give an overview of the observations used in this work. In Section \ref{sec:activity}, we explore a discrepancy in photometric periods and the activity of TRAPPIST-1.  In Section \ref{sec:atmos}, we discuss the predicted effect of stellar activity on the potential atmospheres of the TRAPPIST-1 planets. Finally, in Section \ref{sec:disc}, we discuss our findings.

\section{Observations} \label{sec:obs}

In this study, we use light curves from the \emph{Spitzer} Space Telescope, as well as from the ground-based TRAPPIST-North, TRAPPIST-South, and Liverpool Telescopes \citep{gil17}.  The \emph{Spitzer} observations were taken nearly continuously between 2016 September 19 -- October 10 with a cadence of 2.2 minutes. We also use data from the ground-based TRAPPIST-North, TRAPPIST-South, and Liverpool light curves that coincide with these \emph{Spitzer} observations.  The details of the observations are included in the Methods section of \citet{gil17}.  Here, we are interested only in the starspots, so we remove the signature of the planetary transits and flares, simply by removing the data points from the light curves.  

We also used the raw, 29.4-minute long-cadence \emph{K2} Campaign 12 light curve (2016 December 15 -- 2017 March 4) released by the \emph{K2} Guest Observer Office just after the data were downloaded from the satellite.   We used the publicly-accessible light curve extracted by \citet[][see their work for details of the light curve and extraction]{lug17}.  As with the \emph{Spitzer} observations, we removed the signatures of the planetary transits and flares.  The \citet{lug17} \emph{K2} light curve has a trend that increases the flux over the duration of the observations, which we removed by smoothing the light curve over 10 days and dividing out the resultant trend.

\section{Activity} \label{sec:activity}

\subsection{Photometric Periods} \label{sec:period}

Photometric observations have revealed TRAPPIST-1 to have rotation periods of $P_\mathrm{rot} = 1.40$ days \citep[TRAPPIST-South;][]{gil16} and $P_\mathrm{rot} = 3.30$ days \citep[\emph{K2};][]{lug17,vid17}.  \citet{vid17} showed that the 1.40-day period is not detected in their period search.  The 1.40-day period is, however, consistent with the period determined from the stellar radius \citep[$R = 0.117 \ R_\odot$,][]{gil16}, assumed inclination $i = 90^\circ$, and $v \sin i$ 
\citep[$6.0 \pm 2.0$ km s$^{-1}$,][]{rei10} yielding $P_\mathrm{rot} = 0.99^{-0.25}_{+0.49}$ days, which is also not present in the \emph{K2} light curve.   

The \emph{Spitzer}  and  \emph{K2} light curves were both observed with high cadence and nearly continuously.  The light curves have only a 65-day gap between the data sets.  While the \citet{lug17} and \citet{vid17} period searches of the \emph{K2} light curve are robust, we repeat the efforts to ensure consistency with our treatment of the \emph{Spitzer} light curve. 

\begin{figure}
\begin{center}
\includegraphics[angle=90,scale=.35]{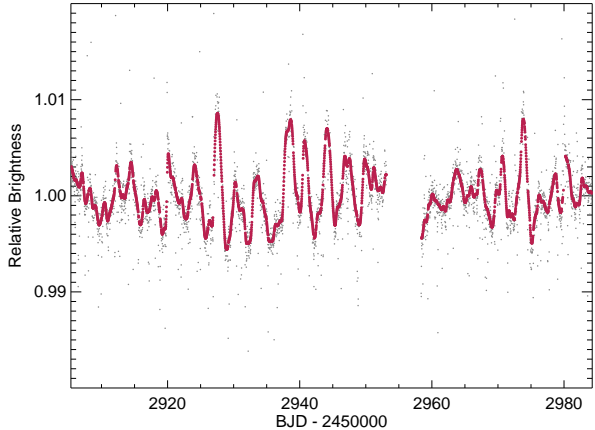}
  \caption{\emph{K2} Campaign 12 light curve OF TRAPPIST-1.  The individual, light gray points are the \emph{K2} observations provided by \citet{lug17} smoothed over 10 days.  The slightly larger red points plotted over the gray represent the 8-hour smoothed \emph{K2} light curve (with eclipses and flares removed) used in our L-S period search.  }
  \label{K2lc}
  \end{center}
\end{figure}

\begin{figure}
\begin{center}
\includegraphics[angle=90,scale=.35]{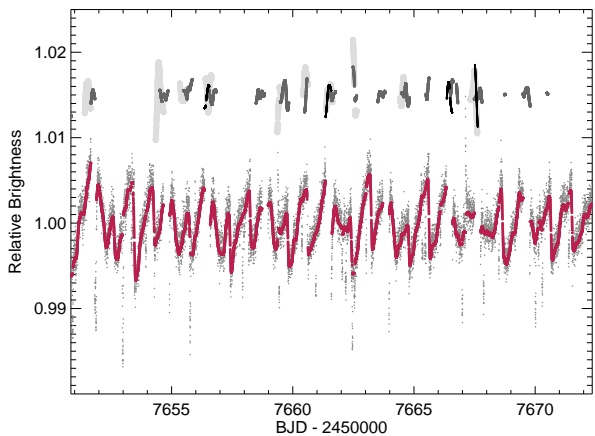}
  \caption{\emph{Spitzer} and ground-based TRAPPIST-North, TRAPPIST-South, and Liverpool light curves of TRAPPIST-1 \citep{gil17}.  The individual, light gray points are the \emph{Spitzer} light curve.  The slightly larger red points plotted over the gray points represent 1.6-hour smoothed \emph{Spitzer} light curve (with eclipses and flares removed).  Offset above the \emph{Spitzer} light curve, the gray lines are the TRAPPIST-North (thickest, darkest gray), TRAPPIST-South (medium thickness, medium gray), and Liverpool Telescope (thinnest, lightest gray) light curves smoothed over 1.6 hours.  }
  \label{SpTNSL}
  \end{center}
\end{figure}

In order to perform a period search on the observations, we prepared the \emph{K2} light curve by smoothing over eight hours ($\sim 16$ data points) to reduce noise (see Figure \ref{K2lc}).  We smoothed the higher-cadence \emph{Spitzer} light curve over 1.6 hours for consistency in the approximate number of data points used in the smoothing length (see Figure \ref{SpTNSL}).

\begin{figure*}
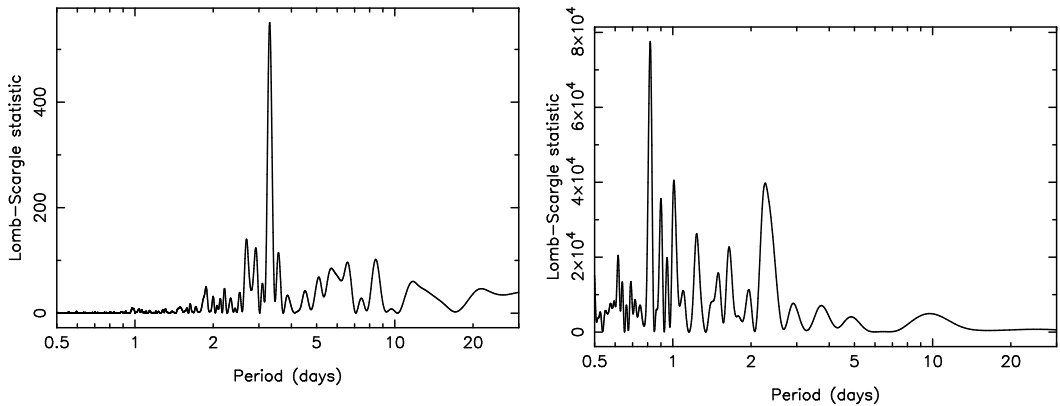

 \centering
\begin{subfigure}
\centering
\includegraphics[angle=270,scale=.31]{t1_k2_pgram.eps} 
\end{subfigure}
\begin{subfigure}
\centering
\includegraphics[angle=270,scale=.31]{t1_spitzer_1_6.eps}
 \end{subfigure}
  \caption{Weighted L-S periodograms.  Left:  For \emph{K2} observations from 2016 December 15 -- 2017 March 4, the L-S periodogram shows the most significant peak at $P_1 = 3.30$ days, which is consistent with that found by \citet{lug17} and \citet{vid17}.  Right:  For \emph{Spitzer} observations from 2016 September 19 -- October 10, the L-S periodogram shows the most significant peak at $P_1 = 0.819$ days.   }
  \label{LSplots}
\end{figure*}

For our period-search analysis, we performed a weighted Lomb-Scargle (L-S) Fourier analysis \citep[as in][]{roe16a}.  The L-S periodograms for the \emph{Spitzer} and \emph{K2} data are found in Figure \ref{LSplots}.  For the \emph{K2} light curve, we confirm that the most prominent period occurs at $3.30 \pm 0.08$ days \citep[consistent with those of][]{lug17,vid17}.  For the \emph{Spitzer} light curve, we found that the strongest period occurred at $0.819 \pm 0.015$ days, a value consistent with the measured $v \sin i$ of \citet{rei10}.  Because of this discrepancy, we divided the \emph{K2} light curve into three portions and performed the L-S Fourier analysis on each to look for evolution within the data set.  We found periods of 3.22, 3.30, and 3.32 days for each third of the data, all of which are consistent within  the period determined using the whole \emph{K2} light curve.

The \emph{Spitzer} photometric period is approximately one-fourth the \emph{K2} photometric period. This strongly suggests that the surface evolved significantly between the two data sets.  It is possible that during the \emph{Spitzer} observations, TRAPPIST-1 presented four particularly dark regions (starspots) nearly equally-spaced across the surface in longitude.  These surface features then evolved over time into what was observed by \emph{K2}---a spotted surface particularly dominated by a single region strongly indicating the rotation period.  Because the TRAPPIST-South period reported by \citet{gil16} is about half of the \emph{K2} period, we suggest that during those observations the star is rotating at the \emph{K2} period and had two regions that were particularly spotted on nearly opposite sides of the star.  This is consistent with a warning given by \citet{irw11} stating that light curves of stars with active starspot regions evenly spaced in longitude could lead to misidentifying the rotation period.

In order to investigate the possibility of the \emph{Spitzer} and \emph{K2} probing the stellar surface of  TRAPPIST-1 differently due to the wavelengths of observation (effective wavelengths of $4.5 \ \mu$m and 600 nm, respectively), we considered the contrast of the starspots to the photosphere.     TRAPPIST-1 is an M8 star with an effective temperature given as $2559 \pm 50$ K \citep{gil17}.  If we extend the trend of \citet[][her Figure 7]{ber05} to cooler stars, the temperature difference between the photosphere and starspots is likely to be $\sim 300$ K.  We note that this trend may not be applicable to fully convective stars.  It is then possible that starspots for such stars could have a larger temperature differential, making the contrast ratios we present here upper limits.  Using the PHOENIX model spectra \citep{dot08} for estimated photospheric and spot temperatures of $T_\mathrm{phot} = 2500$ K and $T_\mathrm{spot} = 2200$ K with surface gravity $\log g = 3.0$ and iron abundance [Fe/H] = 0.0, we find spot-to-photosphere flux ratios of $\left(f_\mathrm{spot}/f_\mathrm{phot} \right)_{Spitzer} = 0.698$ and $\left(f_\mathrm{spot}/f_\mathrm{phot} \right)_{Kepler} = 0.309$.  While the spots will appear fainter in the \emph{Spitzer} bandpass, they will still be detectable.  

Because of the differing effective wavelengths, it is possible that \emph{K2} and \emph{Spitzer} probe different depths; however, we note that a late-M dwarf like TRAPPIST-1 is fully convective and should not have multiple internal layers \citep{lim58,cha97}.  While it seems unlikely that different depths of TRAPPIST-1 will have different rotation periods, the only way to determine if this phenomenon of different periods is the result observing in different wavelengths is to observe simultaneously in different photometric bandpasses.  Fortunately, the \emph{Spitzer} observations coincided with ground-based observations from TRAPPIST-North, TRAPPIST-South, and the 2-m Liverpool Telescope (effective wavelengths of 885 nm, 885 nm, and 900 nm, respectively; see Figure \ref{SpTNSL}). We smoothed the TRAPPIST-North, TRAPPIST-South, and Liverpool Telescope light curves over 1.6 hours and compared those to the smoothed \emph{Spitzer} light curve.  We find qualitatively good agreement between the light curves and note that there is not enough data available from the ground-based observations to reliably perform a period search.  There are regions when the near-infrared, ground-based light curves deviate from the far-infrared, space-based \emph{Spitzer} light curve, but the trends are overall very similar suggesting that observations in different wavelengths are not probing depths that rotate at different rates.   
Because we do not expect this difference in flux ratios or bandpasses observing different spot patterns to have contributed to different period detections, only the changing stellar surface could account for the different photometric period measurements.

\subsection{Rossby Number and Activity} \label{sec:rossby}

The Rossby number, $R_o$, is used to quantify stellar activity and is defined as $R_o \equiv P_\mathrm{rot}/\tau_\mathrm{C}$, where $P_\mathrm{rot}$ is the rotational period and $\tau_\mathrm{C}$ is the convective turnover time ($\tau_\mathrm{C}$ depends only upon spectral type). Studies of various activity measures (H$\alpha$, Ca\begin{footnotesize}II\end{footnotesize}H \& K, X-ray emission, Zeeman-splitting of FeH absorption lines, flare strength) have shown that stars with Rossby numbers of $R_o \lesssim 0.1$ (or $\log R_o \lesssim -1$) have saturated activity; that is, faster rotation will not increase the  observed amount of activity \citep[e.g.,][]{mam08,rei09,wri11,jef11,dou14,wri16,dav16,new17}.

\citet{noy84} first showed $\tau_\mathrm{C}$ increases with decreasing stellar temperature.  For stars with $B-V > 0.9$, \citet{noy84} assigned a constant value, which would be applicable to a typical M8 star \citep[$B-V = 2.2$; ][]{pec13}.  However, further studies have shown that $\tau_\mathrm{C}$ does not become constant for cool stars \citep{kir07,rei10,jef11}.  \citet{rei10} calculated $\log R_o = -1.91$ for TRAPPIST-1 using their $v \sin i$ and estimated stellar radius of $R = 0.1 \ R_\odot$.  They also assume that $\tau_\mathrm{C} = 70$ days for a late M dwarf \citep{gil86,saa01}.  If we update this $R_o$ using the \emph{K2} period and the assumed $\tau_\mathrm{C}$, $\log R_o = -1.33$ ($R_o = 0.047$).  This value is in the saturated regime of the activity-rotation relation.  
Other convective turnover time estimates include $\tau_\mathrm{C} \approx 100$ days for a star with $M \sim 0.08 \ M_\odot$ \citep{kir07} and 560 days using Equation 3 of \citet{jef11}.  Increasing $\tau_\mathrm{C}$ to either of these values or reducing $P_\mathrm{rot}$ to the \emph{Spitzer} or TRAPPIST-S photometric periods will only shift the $R_o$ of TRAPPIST-1 to smaller values, further into the saturated regime.

\begin{figure}
\begin{center}
\includegraphics[angle=90,scale=.35]{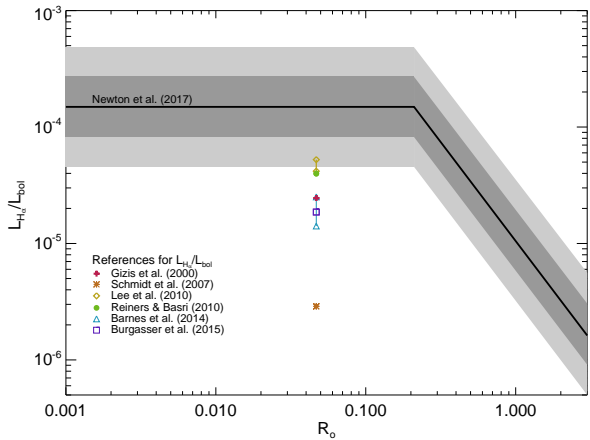}
  \caption{Activity-rotation diagram with literature values of $L_{\mathrm{H}_\alpha}/L_\mathrm{bol}$ for TRAPPIST-1.  The solid lines show the saturated and unsaturated trends found by \citet{new17} with their 1- and 2-$\sigma$ errors assuming intrinsic scatter (darker and lighter gray regions, respectively).  We plot the literature values of $L_{\mathrm{H}_\alpha}/L_\mathrm{bol}$ at the $R_o$ that assumes $\tau_\mathrm{C} = 70$ days and the \emph{K2} rotation period of $3.30$ days for $R_o = 0.047$ ($\log R_o = -1.33$). Each data point is represented as a different character; those connected by a solid line represent the minimum and maximum $L_{\mathrm{H}_\alpha}/L_\mathrm{bol}$ reported by the reference.  }
  \label{actrot}
  \end{center}
\end{figure}

The location of TRAPPIST-1 in the saturated regime is somewhat contradicted by studies that measured the H$\alpha$ luminosity of the star.  In Figure \ref{actrot}, we show literature values of $L_{\mathrm{H}_\alpha}/L_\mathrm{bol}$ for TRAPPIST-1 at the $R_o$ value using the \emph{K2} rotation period and $\tau_\mathrm{C} = 70$ days.  These observations of H$\alpha$ luminosity are below the best-fit saturated value of \citet{new17} for a population of nearby M dwarfs.  However, the highest observed activity levels for TRAPPIST-1 are reasonably consistent with the \citet{new17} $2\sigma$ lower limit of the activity-rotation relationship (see their Figure 7).  

In a detailed study of various age measures, 

\noindent{}\citet{bur17} suggest that TRAPPIST-1 is a $7.6\pm2.2$-Gyr star that still exhibits activity, but only weakly (consistent with \citealt{fil15} and \citealt{lug17}, which estimate ages of 0.5--10 Gyr and 3--8 Gyr, respectively, but in opposition to \citealt{bou17}, which suggests the system is young based on X-ray and UV observations).  While \citet{vid17} describe energetic flaring events, \citet{bur17} emphasize that the events are not frequent enough for an active star.  While the age of late M dwarfs cannot constrain the rotation period, a higher age estimate for TRAPPIST-1 begins to explain the observed low $L_\mathrm{H\alpha}$ values in comparison to the \citet{new17} activity-rotation relationship.

\section{Implications for the Planetary Atmospheres} \label{sec:atmos}

It has been recently noted that the activity of the TRAPPIST-1 host star can have a profound effect on the ability of the detected planets to maintain significant atmospheres \citep{lam03,lam08,san11,don17,gar17,vid17}. In particular, the strong, extreme ultraviolet (XUV) component of the incident flux received by the planets during periods of significant activity can result in ionization of the upper atmosphere of a planet \citep{zen10,vid13}. The extent of the atmospheric loss for a given XUV flux depends on numerous planetary parameters, such as mass, radius, magnetic field, composition, and atmospheric scale height. With our Solar System, there has been observed a relationship between the planetary escape velocity and the flux received by the planet, referred to as the ``cosmic shoreline'' \citep{zah13}. A quantification of this effect has been provided by \citet{zah17}, wherein they apply the methodology to the Proxima Centauri system and propose that the shoreline scales with the incident XUV flux ($I_\mathrm{XUV}$) and escape velocity ($v_\mathrm{esc}$) as $I_\mathrm{XUV} \propto v_\mathrm{esc}^4$. This is especially important for the TRAPPIST-1 system given the age determined by \citet{bur17}, the activity index described in Section~\ref{sec:rossby}, and the XUV fluxes that have been measured for the star \citep{bol17,bou17,whe17}.  As noted earlier, the combination of these stellar attributes ensures that the TRAPPIST-1 planets will have been subjected to extended periods of activity-related atmospheric erosion. As such, the possible atmospheric erosion effects will need to be accounted for in consideration of the habitability potential of the planets.

\begin{figure}
\begin{center}
\includegraphics[angle=270,scale=.35]{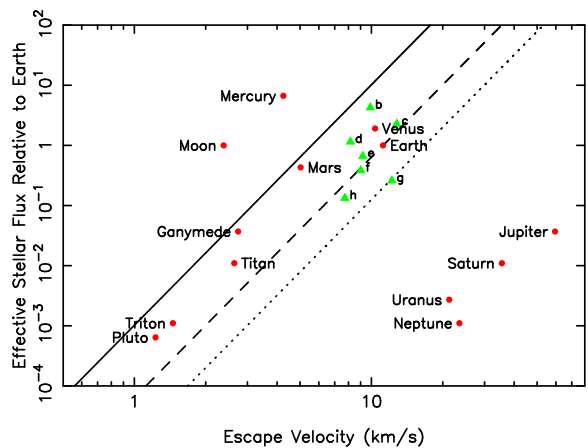}
  \caption{The effective stellar flux incident upon the planet as a function of the calcated escape velocity (km/s) for Solar System object (red circles) and the TRAPPIST-1 planets (green triangles). Two additional lines for the cosmic shoreline downweight the escape velocity by a factor of 2 (dashed) and 4 (dotted) to incorporate the effect of larger XUV flux compared with solar values.}
  \label{csfig}
  \end{center}
\end{figure}

We calculated the incident flux and escape velocities for the TRAPPIST-1 planets based on the planetary and stellar properties provided by \citet{gil17}, where the bolometric luminosity is the result of spectral energy distribution analysis by \citet{fil15}. These calculated values are shown in Table~\ref{cstab}, and are depicted in Figure~\ref{csfig} along with the incident flux and escape velocities for various Solar System objects. The cosmic shoreline based upon the estimates by \citet{zah17} is shown as a solid line, where the Solar System objects to the left of the line tend to not have atmospheres and objects to the right tend to have (albeit sometimes tenuous) atmospheres. Note that the incident flux shown in the plot represents the total (bolometric) flux received at the distance of the planets from the host star.  The derivation of the cosmic shoreline power law by \citet{zah17} utilizes the scaling relationships of \citet{lam09} that relate EUV and XUV excess to stellar age and spectral type. Thus, our adoption of the bolometric flux includes the atmospheric erosion effects of the intrinsic EUV and XUV radiation as a subset of the total flux. The XUV flux related to magnetic activity, as described in this paper, will add an additional component to the atmospheric eroding flux effects.  Increasing the XUV component has the effect of down-weighting the influence of the escape velocity for retaining the atmosphere. This means that low-mass stars with a higher relative rate of XUV emission will have a substantially greater atmospheric erosion effect on their hosted planets. In Figure~\ref{csfig}, we include two additional lines for the cosmic shoreline that reduce the escape velocity effect by factors of 2 (dashed) and 4 (dotted). These lines thus represent stars for which the XUV flux is 16 and 256 times higher than the solar flux. In the case of the former, this indicates that only the g planet has a reasonable chance of retaining a planetary atmosphere, whereas in the case of the latter, all of the planets are likely to be barren. These shifts of the cosmic shoreline are consistent with the XUV flux measurements conducted by \citet{whe17} of TRAPPIST-1, which indicated that the incident flux received by the planets may be several orders of magnitude larger than that received by the Earth. Such incident flux would result in atmosphere and ocean removal over Gyr timescales \citep{bol17}. Clearly, scaling from the solar system model of the cosmic shoreline is prone to unconsidered effects of the stellar and planetary parameters. These unconsidered effects include, but are not limited to, the planetary magnetic field, atmospheric mean molecular weight, stellar age, and the spectral energy distribution (SED). For example, a particularly strong intrinsic dipole magnetic field associated with a planet may provide shielding of the upper atmosphere against erosion caused by XUV radiation, and the age of the star is directly related to the sustained period during which such atmospheric erosion has been able to occur. The concept of the cosmic shoreline as it is applied here is thus intended to serve as a first order effect on the sustainability of the planetary atmospheres.

\begin{deluxetable}{lcc}
\tablecolumns{3}
\tabletypesize{\scriptsize}
  \tablecaption{\label{cstab} TRAPPIST-1 incident (bolometric) flux and escape velocities}
  \tablewidth{0pc}
  \tablehead{
    \colhead{Planet} &
    \colhead{Incident flux} &
    \colhead{Escape velocity} \\
	\colhead{} &
    \colhead{(Relative to Earth)} &
    \colhead{(km/s)}
}
  \startdata
  b & 4.27 & 9.9 \\
  c & 2.28 & 12.8 \\
  d & 1.15 & 8.1 \\
  e & 0.66 & 9.2 \\
  f & 0.38 & 9.0 \\
  g & 0.26 & 12.2 \\
  h & 0.13 & 7.7 \\
  \enddata
\end{deluxetable}

\section{Discussion} \label{sec:disc}

TRAPPIST-1 is a late-M dwarf that is of particular interest due to its many planets, some of which fall in the star's HZ.  Here, we have highlighted the discrepancies of the photometric periods determined by photometric period searches of the independent \emph{Spitzer} and \emph{K2} light curves.  We have added to the list of potential periods obtained via photometry with our analysis of the 20-day \emph{Spitzer} light curve indicating a period one-fourth of that seen in the \emph{K2} observations.  While our results ultimately support the rotation period determined by analyzing the \emph{K2} light curve, it is clear that M dwarfs can host very spotted surfaces that will mislead photometric period searches, suggesting shorter rotation periods.  It is still possible that the \emph{K2} rotation period could be inaccurate as the value does not agree with the observed $v \sin i$ \citep{rei10} and the surface of TRAPPIST-1 rapidly evolves.    We assert that determining rotation periods photometrically is not  reliable for this and similar stars.  In fact, because the surface of TRAPPIST-1 is likely a complicated mix of dark and bright patches rapidly evolving as the star rotates, it might be more reliable to determine rotation periods using the projected rotational velocity ($v \sin i$) of the star, which is determined with more than one spectral line, and ideally more than one spectrum \citep{fek97}.  For example, the $v \sin i$ value of TRAPPIST-1 was determined using the many absorption features of the FeH band near 1 $\mu$m \citep{rei10}.  

Light curves and also perhaps spectra will not allow us to understand the activity of such a rapidly-evolving system.  Rapid evolution of surface features creates significant difficulties ruling out the use of imaging techniques used on more massive or more evolved stars \citep[e.g., Doppler and interferometric imaging, see][for example]{kor09,roe16b}. 

An important implication of the stellar activity combined with the low mass of the star is the incident high-energy flux on the planets in the system. The so-called ``cosmic shoreline'' that appears to draw a first-order relationship between incident XUV flux and planetary escape velocities in our own Solar System further emphasizes the importance of this issue. Our scaling of the cosmic shoreline to larger XUV fluxes lends credence to quantitative estimates that the atmospheres of at least the inner planets have probably been largely eroded over the known lifetime of the star.

Sporadic photometry of the stars at the cool end of the Hertzsprung-Russel diagram does not completely reveal their complicated nature.  A high-cadence, longterm photometric study of a large number of late-M dwarfs could be used for understanding how these active stellar surfaces evolve.  However, M dwarfs are faint objects \citep[e.g., $V = 11.13$ for Proxima Centauri;][]{jao14} often making them difficult targets, even for visible-wavelength space satellites like \emph{Kepler}.  Such a study would ideally focus on infrared wavelengths where M dwarfs are brighter.  Additionally, detailed, individual studies of the activity of TRAPPIST-1 and similar stars will aid our understanding of the habitability of terrestrial planets, for example, the age-determination study of \citet{bur17}.  Multiwavelength studies can further quantify the activity of late-M dwarfs though measuring different activity measures (e.g., XUV, Ca\begin{footnotesize}II\end{footnotesize}H\&K, H$\alpha$, etc.) and can be compared to the flare and starspot evolution to find correlations. Ultimately, such detailed studies of stellar activity and rotation rates will be critical in the correct interpretation of current and future exoplanet detections around cool stars.

\section*{Acknowledgements}

We thank M.\ Gillon for providing us with the photometry used in \citet{gil16,gil17}.  We acknowledge the use of the raw \emph{K2} light curve extracted by \citet{lug17}, which they have made publicly available.  We also thank S. Ciceri, B.T. Montet, and the anonymous referee for their useful discussions and comments on this work.  

This paper includes data collected by the \emph{Kepler} mission. Funding for the \emph{Kepler} mission is provided by the NASA Science Mission directorate.  This work is based in part on archival data obtained with the \emph{Spitzer} Space Telescope, which is operated by the Jet Propulsion Laboratory, California Institute of Technology under a contract with NASA.
This research has also made use of the SIMBAD database,
operated at CDS, Strasbourg, France and the Habitable Zone Gallery at hzgallery.org.

\end{document}